\documentclass[pdftex,10pt,conference]{IEEEtran}

\usepackage[pdftex]{graphicx}
\usepackage{curves,epic,latexsym,amsmath,amssymb,array,cite,ifthen}

\newcommand{\ignore}[1]{}

\newcommand{\visual}{}

\newcommand{\Fig}[1]{Fig.~\ref{#1}}

\newcommand{\eqdef}{\stackrel{\scriptscriptstyle\bigtriangleup}{=} }
\newcommand{\eqm}[1]{#1}    % for IEEEtran
\newcommand{\smpl}[2]{#1^{(#2)}}

\newcommand{\E}{\operatorname{E}}

\newcommand{\argmax}{\operatornamewithlimits{argmax}}

\newcommand{\fA}{f_{\mathrm{A}}}
\newcommand{\fB}{f_{\mathrm{B}}}
\newcommand{\pB}{p_{\mathrm{B}}}

\newcounter{examplecntr}
{\begin{trivlist}\small\item[]\refstepcounter{examplecntr}%
 {\bfseries Example~\theexamplecntr%
  \ifthenelse{\equal{#1}{}}{}{ (#1)}.
}}%
{\end{trivlist}}

\newcounter{theoremcntr}
{\begin{trivlist}\item[]\refstepcounter{theoremcntr}%
{\bfseries Theorem~\thetheoremcntr%
  \ifthenelse{\equal{#1}{}}{}{ (#1)}.
}}%
{\hfill$\Box$\end{trivlist}}

\newcommand{\eproofnegspace}{\\[-1.5\baselineskip]\rule{0em}{0ex}}

\newcommand{\cent}[1]{\makebox(0,0){#1}}
\newcommand{\pos}[2]{\makebox(0,0)[#1]{#2}}

\begin{document}

\newcommand{\toprightnote}{%
\begin{picture}(0,0)(0,0)
\put(152,15){\normalsize\texttt{ISIT 2005}}
\end{picture}%
}

\title{\toprightnote%
Expectation Maximization as Message Passing}

\author{\authorblockN{Justin Dauwels, Sascha Korl, and Hans-Andrea Loeliger}
\authorblockA{Dept.\ of Information Technology and Electrical Engineering, 
  ETH, CH-8092 Z\"urich, Switzerland.% 
  %Email: \texttt{loeliger@isi.ee.ethz.ch}
}}

\maketitle

\begin{abstract}
Based on prior work by Eckford, 
it is shown how expectation maximization (EM) may be viewed, and used, 
as a message passing algorithm in factor graphs. 
%it is shown how expectation maximization (EM) can be used 
%as a general tool for estimation in factor graphs. 
%From this perspective, the generalization of EM to 
%parameter vectors (or processes) with nontrivial a priori models 
%becomes obvious.
\end{abstract}

\section{Introduction}
\label{sec:Introduction}

Graphical models \cite{JoSe:gm} such as 
factor graphs \cite{KFL:fg2000} are tools both for system modeling 
and for the development of algorithms for detection and estimation, 
cf.\ \cite{Lg:ifg2004}, \cite{ISCCSP2004c}. 
In addition to the basic sum-product and max-product (or min-sum) algorithms,
which dominate coding applications, signal processing techniques 
including LMMSE/Kalman filtering, gradient algorithms, 
and particle filters can be naturally viewed and used as message passing 
in factor graphs \cite{Lg:ifg2004}, \cite{ISCCSP2004c}. 

Expectation maximization (EM) \cite{DLR:EM1977} \cite{StSe:cmem2004} 
has also become a standard technique for parameter estimation in graphical models 
\cite{Lau:EMgm1995} \cite{Gha:ul2004}. 
In particular, Eckford showed how EM can be viewed, and used, 
as a technique for breaking cycles in factor graphs 
\cite{EcPa:imdgm2000c}, \cite{Eck:ceem2004c}. 
However, it is not obvious if and how EM can be described 
as a \emph{message passing} algorithm with local message update rules. 

%In fact, the problem was solved already in 2000 by Eckford, who showed
%that EM can indeed be naturally used with factor graphs 
%\cite{EcPa:imdgm2000c}, \cite{Eck:ceem2004c}. 
%%Although it is not emphasized in \cite{EcPa:imdgm2000c}, \cite{Eck:ceem2004c}, 
%This discovery not only allows to describe classical EM algorithms  
%(e.g., Baum/Welch parameter estimation for hidden Markov models) 
%by means of factor graphs, but extends the applicability 
%of EM techniques to graphical models beyond classical state space models. 

%However, it appears that Eckford's work has remained largely unnoticed.
%\footnote{Although we have been working with EM techniques for half a year, 
%we only just realized the significance of Eckford's work ourselves.}.
%\footnote{As it often happens, 
%we realized the significance of this prior work only after we had re-invented 
%it ourselves.}. 
%In the present paper, we wish to draw attention to this discovery 
%by giving a concise description 
%(more explicit than in \cite{EcPa:imdgm2000c}, \cite{Eck:ceem2004c}) 
%of EM as message passing in factor graphs. 
%In the present paper, we wish to draw attention to this discovery 
%by showing explicitly how EM may be viewed as message passing in a factor graph. 
%(The message passing interpretation was not made explicit in 
%\cite{EcPa:imdgm2000c}, \cite{Eck:ceem2004c}.) 

In the present paper, we develop EM as a message passing technique. 
The standard ``global'' view of EM is thus replaced by a ``local'' 
message passing view with a new (local) message computation rule 
for continuous variables. 
The new message computation rule can often be used in cases 
where the standard sum-product (integral-product) rule 
%yields intractable (i.e., impractical) integrals. 
yields impractical expressions for the messages. 
%Although we are actively working with several applications, 
%we will here focus on the bare principles and not discuss any applications. 

\section{Review of EM Algorithm}
\label{sec:EM}

We begin by reviewing the expectation maximization (EM) algorithm 
%\cite{DLR:EM1977}, \cite{StSe:cmem2004} 
in a setting which is suitable for the purpose of this paper. 
% in a setting which is slightly more general than the standard setting. 
Suppose we wish to find 
\begin{equation} \label{eqn:thetamax}
\hat\theta_{\mathrm{max}} \eqdef \argmax_{\theta} f(\theta).
\end{equation}
We assume that $f(\theta)$ is the ``marginal'' 
of some real-valued function $f(x,\theta)$:
\begin{equation}
f(\theta) = \int_x f(x,\theta),
\end{equation}
where $\int_x g(x)$ denotes either integration or summation of $g(x)$ 
over the whole range of $x$.
The function $f(x,\theta)$ is assumed to be nonnegative:
\begin{equation}
f(x,\theta) \geq 0 \text{~~~~for all $x$ and all $\theta$}.
\end{equation}
We will also assume that the integral (or the sum) 
$\int_x f(x,\theta) \log f(x,\theta')$ exists for all $\theta$, $\theta'$. 
The EM algorithm attempts to compute (\ref{eqn:thetamax}) as follows:
\begin{enumerate}
\item
Make some initial guess $\smpl{\hat\theta}{0}$.
\item
Expectation step: evaluate
\begin{equation} \label{eqn:EMEstep}
\smpl{f}{k}(\theta) \eqdef \int_x f(x,\smpl{\hat\theta}{k}) \log f(x,\theta).
\end{equation}
\item
Maximization step: compute
\begin{equation} \label{eqn:EMmaxstep}
\smpl{\hat\theta}{k+1} \eqdef \argmax_{\theta} \smpl{f}{k}(\theta).
\end{equation}
\item
Repeat 2--3 until convergence or until the available time is over. 
\end{enumerate}
The main property of the EM algorithm is 
\begin{equation} \label{eqn:EMTheorem}
f(\smpl{\hat\theta}{k+1}) \geq f(\smpl{\hat\theta}{k}).
\end{equation}
For completeness, a proof of (\ref{eqn:EMTheorem}) is given in the appendix.

\section{Message Passing Interpretation}
\label{sec:GenMessPassInterpret}

We now rewrite the EM algorithm in message passing form. 
%so that it can be used as a tool in factor graphs. 
In this section, we will assume a trivial factorization 
\begin{equation} \label{eqn:TrivialFactorization}
f(x,\theta) = \fA(\theta) \fB(x,\theta),
\end{equation}
where $\fA(\theta)$ may be viewed as encoding the a priori information about 
$\Theta$. 
%Nontrivial factor graphs 
More interesting factorizations 
(i.e., models with internal structure) 
will be considered in the next section.

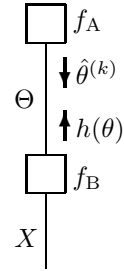
\begin{figure}
\begin{center}
\begin{picture}(5,35)(0,0)
%\put(0,0){\dashbox(20,40){}}
\put(0,30){\framebox(5,5){}}  \put(6,32){$\fA$}
\put(2.5,15){\line(0,1){15}}  \put(1,22.5){\pos{r}{$\Theta$}}
 {\thicklines
  \put(5,28){\vector(0,-1){4}} \put(6.5,24.5){$\smpl{\hat\theta}{k}$}
  \put(5,17){\vector(0,1){4}}  \put(6.5,17.5){$h(\theta)$}
 }
\put(0,10){\framebox(5,5){}}  \put(6,11){$\fB$}
\put(2.5,0){\line(0,1){10}}   \put(1.5,4){\pos{r}{$X$}}
\end{picture}
\caption{\label{fig:TrivialFactorGraph}%
Factor graph corresponding to (\ref{eqn:TrivialFactorization}).}
\end{center}
\end{figure}

We will use Forney-style factor graphs as in \cite{Lg:ifg2004}, 
where edges represent variables and nodes represent factors. 
As in \cite{Lg:ifg2004}, we will use capital letters 
for model variables 
% (random variables or parameters) 
%and lower-case letters for values of such variables. 
and small letters for values of such variables. 
The factor graph of (\ref{eqn:TrivialFactorization}) is shown in 
\Fig{fig:TrivialFactorGraph}. 
In this setup, the EM algorithm amounts to iterative 
recomputation of the following messages:
\begin{description}
\item[Upwards message $h(\theta)$:]
\begin{align}
h(\theta)
&\eqm{=}  \frac{\int_x \fB(x,\smpl{\hat\theta}{k})\log \fB(x,\theta)}{\int_x \fB(x,\smpl{\hat\theta}{k})} 
          \label{eqn:UpwardsMessageTrivialGraph} \\
%&\eqm{=}  \int_x \frac{\fB(x,\smpl{\hat\theta}{k})}{\int_{x'} \fB(x',\smpl{\hat\theta}{k})} \log \fB(x,\theta).
%          \label{eqn:UpwardsMessageTrivialGraphNormedProb} \\
&\eqm{=}  \E_{\pB}\!\left[ \log \fB(X,\theta) \right],
          \label{eqn:UpwardsMessageTrivialGraphExpectation}
\end{align}
%where the expectation in (\ref{eqn:UpwardsMessageTrivialGraphExpectation}) 
%is with respect to the probability distribution 
where $\E_{\pB}$ denotes the expectation with respect to the 
probability distribution 
\begin{equation}
\pB(x|\smpl{\hat\theta}{k}) \eqdef \frac{\fB(x,\smpl{\hat\theta}{k})}{\int_{x'} \fB(x',\smpl{\hat\theta}{k})}.
\end{equation}
\item[Downwards message $\smpl{\hat\theta}{k}$:]
\begin{align}
\smpl{\hat\theta}{k+1} 
&\eqm{=}  \argmax_\theta \left( \log \fA(\theta) + h(\theta) \right)
          \label{eqn:DownwardsMessageTrivialGraph} \\
&\eqm{=}  \argmax_\theta \left( \fA(\theta) \cdot e^{h(\theta)} \right).
          \label{eqn:DownwardsMessageTrivialGraphMultiplicative}
\end{align}
\end{description}

\noindent
The equivalence of this message passing algorithm 
with (\ref{eqn:EMEstep}) and (\ref{eqn:EMmaxstep}) may be seen as follows. 
From (\ref{eqn:EMEstep}) and (\ref{eqn:EMmaxstep}), we have
\begin{align}
\smpl{\hat\theta}{k+1} 
&\eqm{=} \argmax_{\theta} \int_x f(x,\smpl{\hat\theta}{k}) \log f(x,\theta) \\
&\eqm{=} \argmax_\theta \int_x \fA(\smpl{\hat\theta}{k}) \fB(x,\smpl{\hat\theta}{k}) 
               \log\!\big( \fA(\theta) \fB(x,\theta) \big) \\
&\eqm{=} \argmax_\theta \int_x \fB(x,\smpl{\hat\theta}{k}) 
         \Big( \log \fA(\theta) + \log \fB(x,\theta) \Big) \\
%&\eqm{=} \argmax_\theta \left( \int_x \fB(x,\smpl{\hat\theta}{k}) \log \fA(\theta) 
%          + \int_x \fB(x,\smpl{\hat\theta}{k}) \log \fB(x,\theta) \right) \\
&\eqm{=} \argmax_\theta \left( \log \fA(\theta) 
         + \frac{\int_x \fB(x,\smpl{\hat\theta}{k}) \log \fB(x,\theta)}{\int_{x'} \fB(x',\smpl{\hat\theta}{k})}
         \right),
\end{align}
which is equivalent to (\ref{eqn:UpwardsMessageTrivialGraph}) 
and (\ref{eqn:DownwardsMessageTrivialGraph}).

%The following is worth to be pointed out:
%We point out:
Some remarks:
\begin{enumerate}
\item
The computation (\ref{eqn:UpwardsMessageTrivialGraph}) 
% or (\ref{eqn:UpwardsMessageTrivialGraphNormedProb}) 
or (\ref{eqn:UpwardsMessageTrivialGraphExpectation}) 
is \emph{not} an instance of the sum-product algorithm.
%\item
%The computation (\ref{eqn:DownwardsMessageTrivialGraphMultiplicative}) 
%may be viewed as a (trivial) instance of 
%the standard max-product algorithm to compute 
%$\tilde f(\theta) \eqdef \fA(\theta) \cdot e^{h(\theta)}$, 
%followed by the computation of $\argmax_\theta \tilde f(\theta)$. 
\item
The message $h(\theta)$ may be viewed as a ``log-domain'' 
summary of $f_B$. 
In (\ref{eqn:DownwardsMessageTrivialGraphMultiplicative}), 
the corresponding ``probability domain'' summary 
$e^{h(\theta)}$ is consistent with the factor graph interpretation. 
\item
\label{enum:TrivFFGRemarkConstInh}
%Adding a constant to $h(\theta)$ does not affect 
%(\ref{eqn:DownwardsMessageTrivialGraph}) and may therefore be applied freely.
A constant may be added to (or subtracted from) $h(\theta)$ 
without affecting (\ref{eqn:DownwardsMessageTrivialGraph}). 
\item
If $\fA(\theta)$ is a constant, the normalization 
in (\ref{eqn:UpwardsMessageTrivialGraph}) can be omitted. 
More generally, the normalization 
in (\ref{eqn:UpwardsMessageTrivialGraph}) can be omitted 
if $\fA(\theta)$ is constant for all $\theta$ such that $\fA(\theta)\neq 0$. 
However, in contrast to most standard accounts of the EM algorithm, 
we explicitly wish to allow more general functions $\fA$. 
%\item
%Nothing changes if we introduce known observations 
%$y_\mathrm{A}$ and $y_\mathrm{B}$ (i.e., constant arguments) 
%into $f$, $\fA$, and $\fB$ such that (\ref{eqn:TrivialFactorization}) 
%becomes 
%$f(x,y,\theta) = \fA(y_\mathrm{A},\theta) \fB(x,y_\mathrm{B},\theta)$. 
\item
Nothing changes if we introduce a known observation 
(i.e., a constant argument) $y$ into $f$ 
such that (\ref{eqn:TrivialFactorization}) becomes 
$f(x,y,\theta) = \fA(y,\theta) \fB(x,y,\theta)$. 
\end{enumerate}

\begin{figure*}
\begin{center}
\begin{picture}(110,45)(0,0)
%\put(0,0){\dashbox(110,45){}}

\put(0,35){\framebox(110,10){}}  \put(112,35){$\fA$}
\put(0,4){\dashbox(110,16){}}    \put(112,4){$\fB$}
\put(5,10){\framebox(5,5){}}     \put(7.5,7){\cent{$f_0$}}
\put(10,12.5){\line(1,0){15}}    \put(17.5,15){\cent{$X_0$}}
\put(25,10){\framebox(5,5){}}    \put(26,7){\pos{r}{$f_1$}}
\put(27.5,15){\line(0,1){20}}    \put(26.5,27){\pos{r}{$\Theta_1$}}
 {\thicklines
  \put(30,33){\vector(0,-1){4}}  \put(31.5,29.5){$\hat\theta_1$}
  \put(30,22){\vector(0,1){4}}   \put(31.5,22.5){$h_1(\theta_1)$}
 }
\put(27.5,10){\line(0,-1){11}}   \put(29,0){$y_1$}
\put(30,12.5){\line(1,0){15}}    \put(37.5,15){\cent{$X_1$}}
\put(45,10){\framebox(5,5){}}    \put(46,7){\pos{r}{$f_2$}}
\put(47.5,15){\line(0,1){20}}    \put(46.5,27){\pos{r}{$\Theta_2$}}
 {\thicklines
  \put(50,33){\vector(0,-1){4}}  \put(51.5,29.5){$\hat\theta_2$}
  \put(50,22){\vector(0,1){4}}   \put(51.5,22.5){$h_2(\theta_2)$}
 }
\put(47.5,10){\line(0,-1){11}}   \put(49,0){$y_2$}
\put(50,12.5){\line(1,0){10}}    \put(57.5,15){\cent{$X_2$}}
\put(70,27){\cent{$\ldots$}}
\put(70,10){\cent{$\ldots$}}
\put(80,12.5){\line(1,0){10}}    \put(82.5,15){\cent{$X_{n-1}$}}
\put(90,10){\framebox(5,5){}}    \put(91,7){\pos{r}{$f_n$}}
\put(92.5,15){\line(0,1){20}}    \put(91.5,27){\pos{r}{$\Theta_n$}}
 {\thicklines
  \put(95,33){\vector(0,-1){4}}  \put(96.5,29.5){$\hat\theta_n$}
  \put(95,22){\vector(0,1){4}}   \put(96.5,22.5){$h_n(\theta_n)$}
 }
\put(92.5,10){\line(0,-1){11}}   \put(94,0){$y_n$}
\put(95,12.5){\line(1,0){10}}    \put(102.5,15){\cent{$X_n$}}
\end{picture}
%\vspace{2mm}
\vspace{0mm}
\caption{\label{fig:ChainFactorGraph}%
Factor graph corresponding to (\ref{eqn:ChainFactorization}).}
\end{center}
\end{figure*}
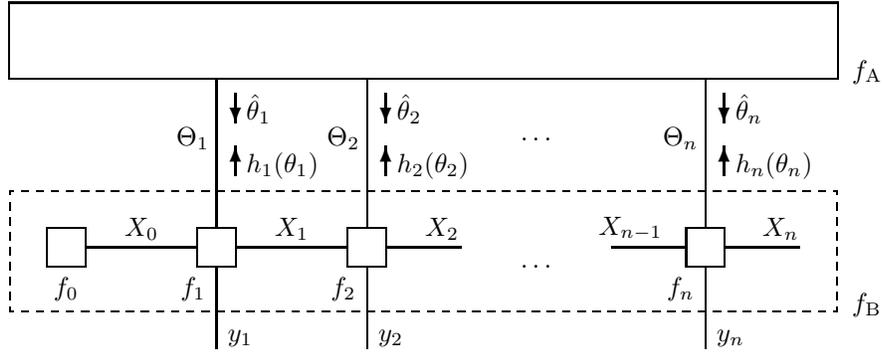

\section{Nontrivial Factor Graphs}
\label{sec:NontrivialFactorGraphs}

The algorithm of the previous section still applies 
if both $\Theta=(\Theta_1,\ldots,\Theta_n)^T$ 
and $X=(X_0,\ldots,X_n)^T$ are vectors. 
However, opportunities to simplify the computations 
may arise if $\fA$ and $\fB$ have ``nice'' factorizations. 
For example, assume that $\fB$ factors as 
\begin{equation}  \label{eqn:ChainFactorization}
\fB(x,y,\theta) = f_0(x_0) f_1(x_0,x_1,y_1,\theta_1)  %f_2(x_1,x_2,y_2,\theta_2) 
    \cdots f_n(x_{n-1},x_n,y_n,\theta_n),
\end{equation}
where $y=(y_1,\ldots,y_n)^T$ is some known (observed) vector. 
Such factorizations arise from classical trellis models and state space models. 
The factor graph corresponding to (\ref{eqn:ChainFactorization}) 
is shown in \Fig{fig:ChainFactorGraph}. 

The upwards message $h(\theta)$ (\ref{eqn:UpwardsMessageTrivialGraphExpectation}) 
splits into a sum with one term for each node in the factor graph:
\begin{align}
h(\theta)
%&\eqm{=}  \E\!\left[ \log\Big( 
%          f_0(x_0) f_1(x_0,x_1,y_1,\theta_1)  %f_2(x_1,x_2,y_2,\theta_2) 
%          \cdots f_n(x_{n-1},x_n,y_n,\theta_n) \Big) \right] \\
&\eqm{=}  \E\!\Big[ \log\Big( 
          f_0(x_0) f_1(x_0,x_1,y_1,\theta_1)  %f_2(x_1,x_2,y_2,\theta_2) 
          \cdots \nonumber \\
&\hspace{5em}  \cdots f_n(x_{n-1},x_n,y_n,\theta_n) \Big) \Big] \\
%&\eqm{=}  \E\!\left[ \log f_0(X_0) \right]
%          + \E\!\left[ \log f_1(X_0,X_1,y_1,\theta_1) \right]
%          + \ldots + \E\!\left[ \log f_n(X_{n-1},X_n,y_n,\theta_n) \right]
&\eqm{=}  \E\!\left[ \log f_0(X_0) \right]
          + \E\!\left[ \log f_1(X_0,X_1,y_1,\theta_1) \right] + \ldots
          \nonumber\\
&\hspace{5em}
          \ldots + \E\!\left[ \log f_n(X_{n-1},X_n,y_n,\theta_n) \right]
          \label{eqn:ChainUpwardsSum}
\end{align}
Each term 
\begin{equation} \label{eqn:UpwardsMessageChainGraphExp}
h_k(\theta_k) \eqdef \E\!\left[ \log f_k(X_{k-1},X_k,y_k,\theta_k) \right] 
\end{equation}
may be viewed as the message out of the corresponding node, 
as indicated in \Fig{fig:ChainFactorGraph}. 
The constant term $\E\!\left[ \log f_0(X_0) \right]$ in (\ref{eqn:ChainUpwardsSum}) 
may be omitted 
%(as noted in the previous section). 
(cf.\ Remark~\ref{enum:TrivFFGRemarkConstInh} in Section~\ref{sec:GenMessPassInterpret}). 
As in (\ref{eqn:UpwardsMessageTrivialGraphExpectation}),
all expectations are with respect to the probability distribution 
$\pB$, which we here denote by $\pB(x|y,\hat\theta)$. 
Note that each term (\ref{eqn:UpwardsMessageChainGraphExp}) 
requires only $\pB(x_{k-1},x_k | y,\hat\theta)$, 
the joint distribution of $X_{k-1}$ and $X_k$:
\begin{equation}  \label{eqn:UpwardsMessageChainGraphPB}
h_k(\theta_k) = \int_{x_{k-1}} \int_{x_k} 
          \pB(x_{k-1},x_k | y,\hat\theta) \log f_k(x_{k-1},x_k,y_k,\theta_k). 
\end{equation}
These joint distributions may be obtained 
by means of the standard sum-product algorithm (belief propagation) 
\cite{KFL:fg2000} \cite{Lg:ifg2004}: 
from elementary factor graph theory, we have 
%\begin{equation}
%\pB(x_{k-1},x_k | y,\hat\theta) 
%\propto f_k(x_{k-1},x_k,y_k,\hat\theta) \,
%        \mu_{X_{k-1}\rightarrow f_k}(x_{k-1}) \, \mu_{X_k\rightarrow f_k}(x_k),
%\end{equation}
\begin{align}
\pB(x_{k-1},x_k | y,\hat\theta) 
&\eqm{\propto} f_k(x_{k-1},x_k,y_k,\hat\theta) \,
        \mu_{X_{k-1}\rightarrow f_k}(x_{k-1}) \,  \nonumber\\
&\hspace{5em}  \cdot \mu_{X_k\rightarrow f_k}(x_k),
\end{align}
where $\mu_{X_{k-1}\rightarrow f_k}$ and $\mu_{X_k\rightarrow f_k}$ 
are the messages of the sum-product algorithm towards the node $f_k$ and where 
``$\propto$'' denotes equality up to a scale factor that does not depend on 
$x_{k-1}, x_k$. 
It follows that 
%\begin{equation} \label{eqn:pBfromMessages}
%\pB(x_{k-1},x_k | y,\hat\theta) 
%= \frac{f_k(x_{k-1},x_k,y_k,\hat\theta) \,
%       \mu_{X_{k-1}\rightarrow f_k}(x_{k-1}) \, \mu_{X_k\rightarrow f_k}(x_k)}
%       {\int_{x_{k-1}} \int_{x_k} 
%        f_k(x_{k-1},x_k,y,\hat\theta) \,
%        \mu_{X_{k-1}\rightarrow f_k}(x_{k-1}) \, \mu_{X_k\rightarrow f_k}(x_k)}.
%\end{equation}
\begin{align} 
\pB(x_{k-1},x_k | y,\hat\theta) 
& \eqm{=} \nonumber\\
&&\hspace{-9em} \frac{f_k(x_{k-1},x_k,y_k,\hat\theta) \,
       \mu_{X_{k-1}\rightarrow f_k}(x_{k-1}) \, \mu_{X_k\rightarrow f_k}(x_k)}
       {\int_{x_{k-1}} \int_{x_k} 
        f_k(x_{k-1},x_k,y,\hat\theta) \,
        \mu_{X_{k-1}\rightarrow f_k}(x_{k-1}) \, \mu_{X_k\rightarrow f_k}(x_k)}.
   \label{eqn:pBfromMessages}
\end{align}
Note that, 
if the sum product messages $\mu_{X_{k-1}\rightarrow f_k}$ and $\mu_{X_k\rightarrow f_k}$ 
are computed without any scaling, then 
the denominator in (\ref{eqn:pBfromMessages}) equals 
%\begin{equation}
%\pB(y_k|\hat\theta) = ???
%\end{equation}
$\pB(y|\hat\theta)$, which is independent of $k$.

%\begin{equation} \label{eqn:UpwardsMessageChainGraphFromSumProduct}
%h_k(\theta_k) = 
%  \int_{x_{k-1}} \int_{x_k} 
%  \frac{f_k(x_{k-1},x_k,y,\hat\theta) \,
%       \mu_{X_{k-1}\rightarrow f_k}(x_{k-1}) \, \mu_{X_k\rightarrow f_k}(x_k)}
%       {\int_{x_{k-1}} \int_{x_k} 
%        f_k(x_{k-1},x_k,y,\hat\theta) \,
%        \mu_{X_{k-1}\rightarrow f_k}(x_{k-1}) \, \mu_{X_k\rightarrow f_k}(x_k)}
%  \log f_k(x_{k-1},x_k,y_k,\theta_k).
%\end{equation}

\begin{figure}
\visual\vspace{5mm}
\begin{center}
\setlength{\unitlength}{0.8mm}
\begin{picture}(100,15)(0,0)
%\put(0,0){\dashbox(100,15){}}
%
\put(0,5){\dashbox(95,15){}}        \put(97,5){$\fA$}
\put(7.5,12.5){\line(0,-1){13.5}}   \put(9,-1){$\Theta_1$}
\put(7.5,12.5){\line(1,0){17.5}}
\put(25,10){\framebox(5,5){$=$}}
\put(27.5,10){\line(0,-1){11}}      \put(29,-1){$\Theta_2$}
\put(30,12.5){\line(1,0){10}}
\put(47.5,10){\cent{\ldots}}
\put(55,12.5){\line(1,0){10}}
\put(65,10){\framebox(5,5){$=$}}
\put(67.5,10){\line(0,-1){11}}      \put(69,-1){$\Theta_{n-1}$}
\put(70,12.5){\line(1,0){17.5}}
\put(87.5,12.5){\line(0,-1){13.5}}  \put(89,-1){$\Theta_n$}
\end{picture}
%\vspace{2mm}
\vspace{0mm}
\caption{\label{fig:FFGParameterEqualsChain}%
Factor graph of $\Theta_1=\Theta_2=\ldots=\Theta_n$.}
\end{center}
\end{figure}
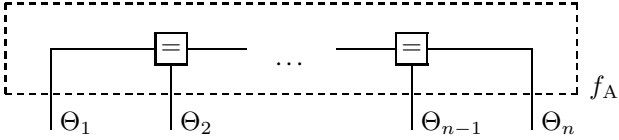

The downwards message $\hat\theta$ (\ref{eqn:DownwardsMessageTrivialGraph}) is 
\begin{align}
(\hat\theta_1,\ldots,\hat\theta_n)^T 
&\eqm{=}  \argmax_{\theta_1,\ldots,\theta_n} 
          \big( \log \fA(\theta) + h_1(\theta_1) + \ldots \nonumber\\
&\hspace{5em} \ldots + h_n(\theta_n) \big) 
          \label{eqn:DownwardsMessageChainGraph} \\
&\eqm{=}  \argmax_{\theta_1,\ldots,\theta_n} 
          \left( \fA(\theta) \cdot e^{h_1(\theta_1)} \cdots e^{h_n(\theta_n)} \right).
          \label{eqn:DownwardsMessageChainGraphMultiplicative}
\end{align}
If $\fA$ has itself a nice factorization,  
then (\ref{eqn:DownwardsMessageChainGraph}) 
or (\ref{eqn:DownwardsMessageChainGraphMultiplicative}) 
may be computed by the standard max-sum or max-product algorithm, 
respectively. 
This applies, in particular, for the standard case 
$\Theta_1=\Theta_2=\ldots=\Theta_n$, 
which is illustrated in \Fig{fig:FFGParameterEqualsChain}.

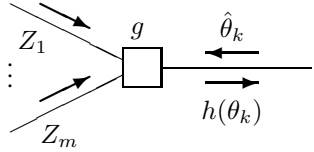
\begin{figure}
\begin{center}
\begin{picture}(45,25)(0,0)
%\put(0,0,){\dashbox(45,25){}}
%
\put(5,21){\line(2,-1){15}}    \put(6,15){$Z_1$}
\put(5,12.5){\cent{$\vdots$}}
\put(5,4){\line(2,1){15}}      \put(9,2.5){$Z_m$}
\put(20,10){\framebox(5,5){}}  \put(21,17){$g$}
\put(25,12.5){\line(1,0){20}}  %\put(35,14){$\Theta$}
\thicklines
\put(9,21){\vector(2,-1){6.5}}
\put(9,8){\vector(2,1){6.5}}
\put(38,14.5){\vector(-1,0){7}}  \put(34.5,18){\cent{$\hat\theta_k$}}
\put(31,10.5){\vector(1,0){7}}   \put(34.5,6.5){\cent{$h(\theta_k)$}}
\end{picture}
\caption{\label{fig:hMessageGeneral} $h$-message out of a generic node.}
\end{center}
\end{figure}

%It should be noted that 
The above derivations do not 
in any essential way depend on the specific example (\ref{eqn:ChainFactorization}). 
In principle, any cut-set of edges in some factor graph 
may be chosen to be the vector $\Theta$. 
However, the resulting subgraphs corresponding to $\fA$ and $\fB$ 
should be cycle-free in order to permit the computation of exact expectations 
($h$-messages) and maximizations ($\hat\theta$-messages). 
The $h$-message out of a generic node $g(z_1,\ldots,z_m,\theta_k)$ 
%(in the $\fB$-part of the graph, 
(cf.\ \Fig{fig:hMessageGeneral}) 
is 
%\begin{equation}
%h(\theta_k) = \int_{z_1}\ldots\int_{z_m} 
%  \frac{g(z_1,\ldots,z_m,\hat\theta_k)\, \mu(z_1)\cdots\mu(z_m)}
%       {\int_{z_1}\ldots\int_{z_m} g(z_1,\ldots,z_m,\hat\theta_k)\, \mu(z_1)\cdots\mu(z_m)}
%  \log g(z_1,\ldots,z_m,\theta_k),
%\end{equation}
\begin{align}
h(\theta_k) 
&\eqm{=} \gamma^{-1}\int_{z_1}\ldots\int_{z_m} 
  g(z_1,\ldots,z_m,\hat\theta_k)\, \mu(z_1)\cdots\mu(z_m)  \nonumber\\
  &\hspace{5em} \cdot \log g(z_1,\ldots,z_m,\theta_k)
  \label{eqn:hMessageGeneral}
\end{align}
with
\begin{equation}
\gamma \eqdef 
 \int_{z_1}\ldots\int_{z_m} g(z_1,\ldots,z_m,\hat\theta_k)\, \mu(z_1)\cdots\mu(z_m)
\end{equation}
and where $\mu(z_1),\ldots,\mu(z_m)$ are the standard sum-product messages. 
Obviously, this message passing rule may also be applied 
to a (sub-) graph with cycles, but then there is no guarantee 
for~(\ref{eqn:EMTheorem}).

%\section{An Example}
%
%
%\begin{verbatim}
%- segmentation?
%- factor node tables!!
%\end{verbatim}
%
%\dotl
%

\section{Conclusion}

Elaborating on prior work by Eckford, 
we have formulated EM in message passing form 
with a new message computation rule (\ref{eqn:hMessageGeneral}). 
In this setting, a main attraction of EM 
is that this message passing rule 
can be evaluated in some cases where the standard sum-product 
or max-product rules yield intractable expressions. 
It is likely that ``local'' use of this message computation rule 
can give good results even in situations where the ``global'' 
conditions required to guarantee (\ref{eqn:EMTheorem}) are not satisfied. 
%In some applications, 
%Local use of such ``expectation messages'' may be attractive 
%even in cases where the ``global'' assumptions for 
%the proof of the main EM property () are not satisfied. 

\section*{Appendix: Proof of (\ref{eqn:EMTheorem})}

The proof is standard (cf.\ \cite{StSe:cmem2004}), 
%but adapted to the slightly nonstandard setting of this paper.
but adapted to the slightly nonstandard setup of Section~\ref{sec:EM}. 
%but adapted to the setup of Section~\ref{sec:EM}. 

\begin{trivlist}\item[]{\bfseries Lemma:}
The function
\begin{equation} \label{eqn:EMLemmaAuxFunc}
\tilde f(\theta,\theta') 
\eqdef f(\theta') + \int_x f(x,\theta') \log \frac{f(x,\theta)}{f(x,\theta')}
\end{equation}
satisfies both
\begin{equation} \label{eqn:EMLemmaAuxFuncIneq}
\tilde f(\theta,\theta') \leq f(\theta)
\end{equation}
and
\begin{equation} \label{eqn:EMLemmaAuxFuncEq}
\tilde f(\theta,\theta) = f(\theta).
\end{equation}
\eproofnegspace\hfill$\Box$
\end{trivlist}
 
\begin{proof}
The equality (\ref{eqn:EMLemmaAuxFuncEq}) is obvious. 
The inequality (\ref{eqn:EMLemmaAuxFuncIneq}) follows from 
eliminating the logarithm in (\ref{eqn:EMLemmaAuxFunc}) by 
the inequality $\log x \leq x-1$ for $x>0$:
\begin{align}
\tilde f(\theta,\theta') 
&\eqm{\leq}  f(\theta') + \int_x f(x,\theta') 
             \left( \frac{f(x,\theta)}{f(x,\theta')} - 1 \right) \\
&\eqm{=}     f(\theta') + \int_x f(x,\theta) - \int_x f(x,\theta') \\
&\eqm{=}     f(\theta).
\end{align}
\eproofnegspace%
\end{proof}

\noindent
To prove (\ref{eqn:EMTheorem}), we first note that 
%(\ref{eqn:EMEstep}) and (\ref{eqn:EMmaxstep}) may be replaced by
(\ref{eqn:EMmaxstep}) is equivalent to
\begin{equation} \label{eqn:maxstepAuxFunc}
\smpl{\hat\theta}{k+1} = \argmax_\theta \tilde f(\theta,\smpl{\hat\theta}{k}).
\end{equation}
We then obtain 
\begin{align}
f(\smpl{\hat\theta}{k}) 
&\eqm{=}    \tilde f(\smpl{\hat\theta}{k}, \smpl{\hat\theta}{k})  \label{eqn:ProofEMLemmaI}\\
&\eqm{\leq} \tilde f(\smpl{\hat\theta}{k+1}, \smpl{\hat\theta}{k}) \label{eqn:ProofEMLemmaII}\\
&\eqm{\leq} f(\smpl{\hat\theta}{k+1}), \label{eqn:ProofEMLemmaIII}
\end{align}
where (\ref{eqn:ProofEMLemmaI}) follows from (\ref{eqn:EMLemmaAuxFuncEq}), 
(\ref{eqn:ProofEMLemmaII}) follows from (\ref{eqn:maxstepAuxFunc}), 
and (\ref{eqn:ProofEMLemmaIII}) follows from (\ref{eqn:EMLemmaAuxFuncIneq}).

\newcommand{\COM}{IEEE Trans.\ Communications}
\newcommand{\IT}{IEEE Trans.\ Information Theory}
\newcommand{\JSAC}{IEEE J.\ Selected Areas in Communications}
\newcommand{\SPMag}{IEEE Signal Proc.\ Mag.}

\end{document}